\documentclass[a4paper,twoside]{article}

\usepackage{epsfig}
\usepackage{subcaption}
\usepackage{calc}
\usepackage{amssymb}
\usepackage{amstext}
\usepackage{amsmath}
\usepackage{amsthm}
\usepackage{multicol}
\usepackage{pslatex}
\usepackage{apalike}
\usepackage{algorithm2e}
\usepackage[bottom]{footmisc}
\usepackage{tabularx}
\usepackage{xcolor}
\usepackage{colortbl}
\usepackage{SCITEPRESS}     

\begin{document}

\title{Modeling HIF-ILK Interaction using Continuous Petri Nets}

\author{\authorname{Viktor Gilin*\sup{1}, Sanne Laauwen*\sup{1}, Yuying Xia*\sup{1}, Noria Yousufi*\sup{1} and Lu Cao\sup{1}}
\affiliation{\sup{1}Leiden Institute of Advanced Computer Science, Leiden University, Einsteinweg 55, Leiden, The Netherlands}
\email{\{v.gilin, a.m.l.laauwen, y.xia, n.s.yousufi\}@umail.leidenuniv.nl, l.cao@liacs.leidenuniv.nl}
\textit{*These authors contributed equally to this study.}}


\abstract{Oxygen concentration in tumor micro-environment is a well-established signal that can induce aggressive cancer behaviour. In particular, low oxygen levels (hypoxia) activate the Hypoxia-Inducible Factor(HIF) pathway which has an array of target systems. One of these systems is Integrin-Linked Kinase (ILK) pathway, which influences key signaling pathways for cell survival, proliferation, and migration. Hence, this paper aimed to explore the interconnection between these two pathways. Using the Petri net modeling tool Snoopy, an established HIF network model was transformed to be a continuous Petri net. Subsequently, the network was expanded to incorporate a feedback element from the ILK pathway to HIF, based on gene expression data. The resulting model conserved the oxygen switch response of the original HIF model and positively amplified HIF's output. Therefore, this model provides a starting point for establishing a system reflecting crucial effect on hypoxia-induced cancer behavior, and could potentially serve as a basis for future drug development. }

\onecolumn \maketitle \normalsize \setcounter{footnote}{0} \vfill

\section{\uppercase{Introduction}}
Breast cancer is a significant global health concern with an increasing trend in prevalence and mortality rates \cite{azamjah2019global}. Hypoxia, a condition of low oxygen within tumor micro-environments, plays a role in its progression. Under low oxygen conditions, the hypoxia-inducible factor-1 (HIF-1) plays a role in regulating certain aggressive cancer traits \cite{lu2010hypoxia,hanahan2022hallmarks}. It mainly activates genes that are critical for tumor growth and survival, including the integrin-linked kinase (ILK) gene \cite{persad2003role}. The focus of this paper is to explore the interplay between HIF-1 and ILK within the context of breast cancer under hypoxic conditions. This was done by transforming an established Petri net for the HIF-1 pathway by Heiner et al. \cite{heiner2010}, to be continuous and to include mechanisms related to the ILK gene specifically.
The Petri net model, depicting the dynamics of the ILK and HIF-1 pathways, was subsequently analyzed using the Snoopy and Charlie software tools \cite{heiner2012snoopy,heiner2015charlie}. This approach allowed for the study of how variations in hypoxia affect the expression of the ILK gene. To enhance the accuracy and refine the network model, experimental results from Western blots were used to check the presence of proteins in these pathways under different oxygen conditions. This provided empirical verification and allowed for further improvements to the model \cite{chou2015novel}. This approach aims to improve the understanding of the pathway dynamics, offering insights that could be used in research towards potential therapeutic strategies. 

\section{\uppercase{Biological Background}}

A critical factor influencing the progression of breast cancers is hypoxia, a condition of low oxygen prevalent within tumor micro-environments due to low diffusion of oxygen to the tissue affected by the tumor \cite{hanahan2022hallmarks}. This condition significantly influences the expression of numerous genes that are critical to the progression of breast cancer.  This paper specifically looks at the MCF-7 cell line, classified as Luminal A cells. Although this sub-type is generally considered less aggressive \cite{orrantia2022subtypes}, there is a substantial amount of research towards hypoxia performed using this specific cell line \cite{chou2015novel,raja2014hypoxia,hsu2016integrin}. This abundance of research provides a solid foundation for modeling, analysis, verification, and understanding the biological interactions. The following subsections specifically discuss findings from this cell line. \\

\subsection{HIF Pathway}

\begin{figure}[h!]
    \centering
    \includegraphics[scale=2.5]{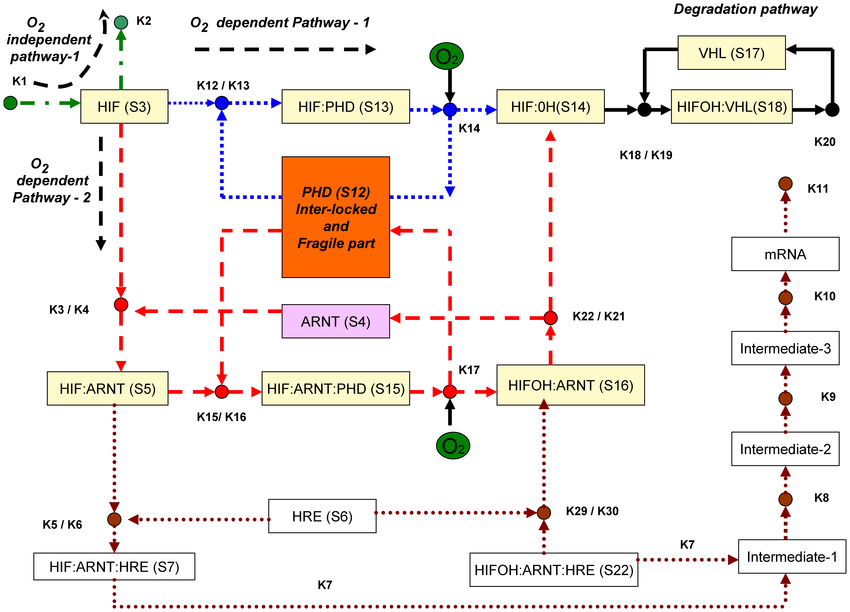}
    \caption{Schematic diagram of hypoxia response network \cite{heiner2010}. Three pathways, given in green (oxygen-independent pathway), blue and red (oxygen-dependent pathways) can degrade HIF transcription factor.}
    \label{fig:HIFdiagram}
\end{figure}

The transcriptional response to hypoxia is primarily mediated by HIF-1, which is a transcription factor regulated by oxygen levels \cite{lin2011differential}. HIFs consist of an oxygen-sensitive \(\alpha\)-subunit and a constitutively expressed \(\beta\)-subunit \cite{muz2015role}. Under normoxic conditions, the HIF-1\(\alpha\) subunit is unstable and hydroxylated by specific prolyl hydroxylases (PHDs). This hydroxylation leads to ubiquitination by the von Hippel-Lindau protein (VHL) and subsequent proteasome-mediated degradation \cite{majmundar2010hypoxia}. In contrast, under hypoxic conditions, the PHDs are inactivated, preventing the hydroxylation and degradation of HIF-1\(\alpha\). This allows HIF-1\(\alpha\) to accumulate.\\
Although HIF-1\(\alpha\) can also degrade through oxygen-independent pathways, these mechanisms are less efficient. Meaning that under low oxygen conditions, HIF-1\(\alpha\) still accumulates. Under hypoxia, HIF-1\(\alpha\) initially binds with high affinity to PHDs, forming a HIF-1\(\alpha\)-PHD complex. However, due to the low oxygen levels, this complex cannot be hydroxylated and thus remains stable. Once the PHDs are saturated, the excess HIF-1\(\alpha\) binds to the \(\beta\)-subunit, also referred to as the ARNT subunit, to form the HIF-1\(\alpha\)-ARNT complex, which actively binds to hypoxia response elements (HREs) in the genome. This complex also associates with PHDs to form a HIF-ARNT-PHD complex, which similarly cannot be hydroxylated due to the low oxygen content. The predominant form under hypoxic conditions is therefore HIF-ARNT, which actively induces HREs. This differential binding behavior of HIF-1\(\alpha\) between PHD and ARNT subunits under varying oxygen levels crucially regulates the cellular adaptation to hypoxia by activating genes that help the cell cope with reduced oxygen availability. Figure \ref{fig:HIFdiagram} provides a schematic representation of the hypoxia response network. In this diagram, HIF1-\(\alpha\) is denoted as HIF, and HIF1-\(\beta\) is referred to as ARNT. The diagram illustrates the protease-mediated degradation pathways of HIF-1\(\alpha\) (shown in black), highlighting both oxygen-dependent (depicted in red and blue) and oxygen-independent mechanisms for degradation (illustrated in green).

\subsection{ILK Gene}
\begin{figure}[h!]
    \centering
    \includegraphics[scale=0.27]{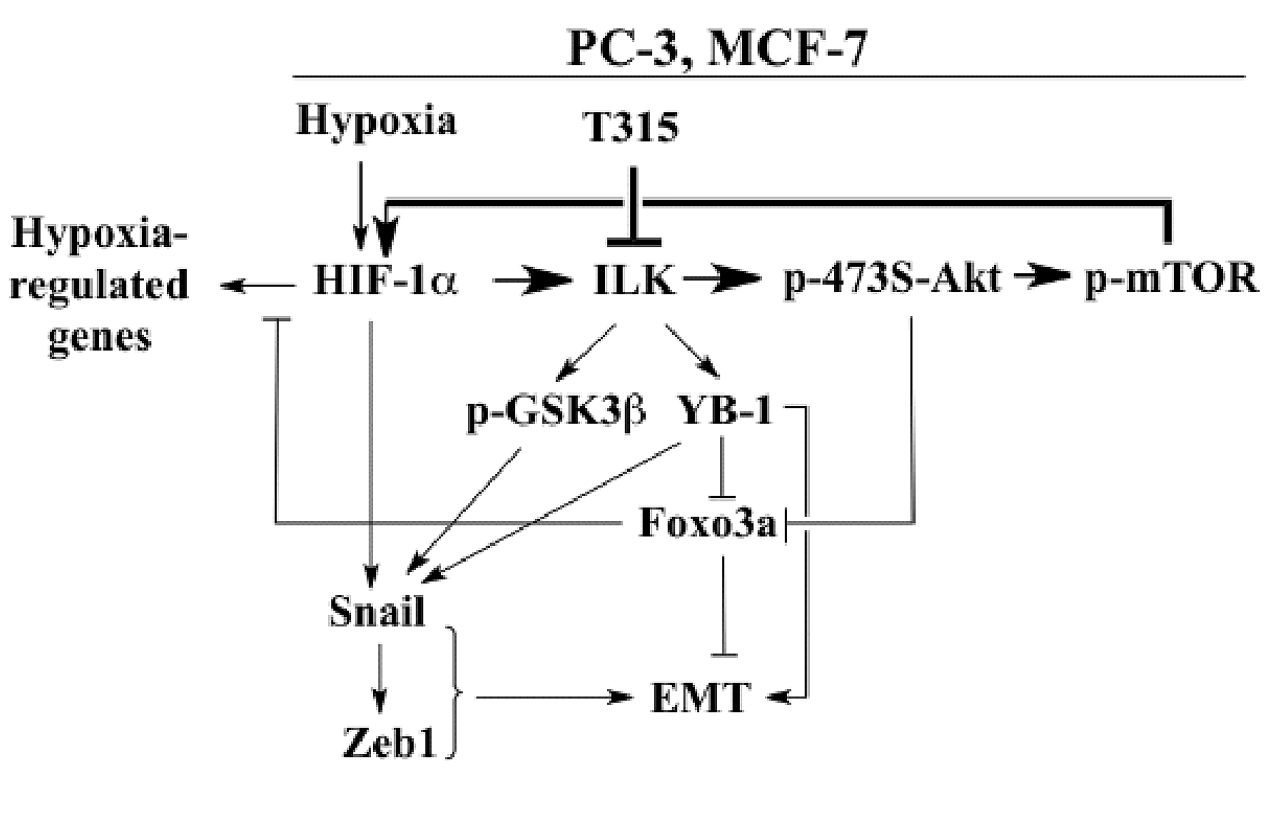}
    \caption{Diagram depicting the mechanisms by which ILK interacts with HIF-1\(\alpha\) expression and the mesenchymal phenotype of cancer cells under hypoxic conditions \cite{chou2015novel}.}
    \label{fig:ILK-signalling}
\end{figure}

The integrin-linked kinase (ILK) gene functions as a serine/threonine protein kinase that interacts with integrins and growth factor receptors to influence key signaling pathways for cell survival, proliferation, and migration \cite{persad2003role,gorska2022integrin}. In conditions of low oxygen, HIF-1\(\alpha\) activates the expression of ILK, which in turn boosts the levels of HIF-1\(\alpha\), stabilizing it during hypoxic stress. This interaction is supported by Akt, which enhances HIF-1\(\alpha\) expression through mTOR-mediated translation, promoting aggressive tumor characteristics and the potential for metastasis \cite{chou2015novel}. Additionally, ILK significantly influences the epithelial-to-mesenchymal transition (EMT), a key process in cancer metastasis, by modulating proteins such as Snail and Twist and by suppressing the tumor suppressor Foxo3a, thereby intensifying HIF-1\(\alpha\) signaling \cite{chou2015novel,emerling2008pten}.
In Figure \ref{fig:ILK-signalling}, a simplified diagram illustrates the mechanisms through which ILK influences HIF-1\(\alpha\) expression under hypoxic conditions. The diagram shows the feedback loop where ILK, through the phosphorylation of Akt at serine 473 (p-473S-Akt), leads to the subsequent activation of mTOR (p-mTOR), which in turn upregulates HIF-1\(\alpha\). The phosphorylation at Ser473 is a critical modification that activates Akt, a kinase involved in various cellular processes such as metabolism, proliferation, and survival. This activation enhances the mTOR pathway, contributing to the upregulation of HIF-1\(\alpha\). Additionally, this pathway regulates downstream factors involved in the EMT, further influencing cellular behavior under hypoxic conditions.

\section{\uppercase{Petri Nets modeling}}
Petri nets are a well-established mathematical formalism initially introduced for modeling concurrent systems in the field of computer science \cite{chaouiya2007petri}. Petri nets represent concurrent, asynchronous and distributed processes in a graphical way, which help in understanding the structure of complex systems. In biology, Petri nets are widely used in modeling and analyzing biological networks such as signaling pathways \cite{terberov2024} and metabolic pathways. A Petri net can be defined mathematically \cite{blatke2011petri2} as a tuple $N = (P,T,F,W,M_0)$. $P$ is the set of places. $T$ is the set of transitions. A relation $F \subseteq (P\times T) \cup (T\times P)$ shows how places and transitions are related. A mapping $W:F \rightarrow \mathbb{N}$ is the weight function that assigns each $arc(x,y)$ connecting a place and a transition an integer $W(x,y)$. The initial marking $M_0: P \rightarrow \mathbb{N}$ assigns each place $p$ its initial token load $M_0(p)$.
Petri nets consist of active components (transitions) and passive components (places), which respectively describe biological activities and entities. By simulating Petri nets, it becomes possible to predict the dynamic behavior of networks without conducting actual experiments. Petri nets allow graphical representation, where places are depicted as circles, transitions as rectangles, and relations as directed arc.\\
Building on the framework of Petri nets, this paper used continuous Petri net to study the dynamics of the ILK and HIF pathways under varying hypoxia conditions. A \textbf{Continuous Petri net} (CPN) is an extension of the basic Petri net. The structure is the same as discrete Petri net. However, the marking of a place is a positive real number, referred to as the token value, which can be interpreted as a concentration \cite{heiner2008petri}. Continuous Petri nets allow for the use of real-time reaction rates and continuous changes in concentrations. In this context, each place represents a variable in the ordinary differential equation (ODE) system, and each transition represents a rate of change in the equations. Reaction rates are implemented through the firing rate function of the transitions. Firing rates can be specified in various ways \cite{herajy2018adaptive}, including mass-action kinetics, Michaelis-Menten kinetics, and Hill kinetics \cite{herajy2018adaptive}. Mass-action kinetics, which describes the firing rates of chemical reactions based on the concentrations of reactants, is used to define the firing rates in the continuous Petri net model presented in this paper. We integrated firing rates from previous simulation studies \cite{kohn_properties_2004} and fine-tuned these firing rates based on experimental data. The model aims to provide a quantitative simulation of how hypoxia levels influence ILK expression through the interaction of the ILK and HIF pathways. In the following subsections, we will discuss the modeling decisions and rationale behind the continuous Petri net used in this case study. 
\subsection{Hypoxia switch response of HIF}
A starting point for our model was the Petri net developed by Heiner et al. \cite{heiner2010}, which has been shown to successfully model HIF's switch response to hypoxia. This model based on the schematic diagram shown in Figure \ref{fig:HIFdiagram}. While Heiner et al. \cite{heiner2010} designed the structure of the network, they only kept it as a discrete Petri net and performed their analysis using different numerical simulation tools. We intend to transform their discrete network into a continuous one. Continuous token values are used to represent the concentrations of the network's elements. This approach allows for a more detailed depiction of the biochemical processes occurring within the system. The Mass-action function is used to implement firing rate values in the continuous transitions. The choice of reaction rates was initially based on those used by Heiner et al. \cite{heiner2010}. The k-value shown in Table \ref{tab:reaction_rates} represents the reaction rate constant obtained by Kohn et al. \cite{kohn_properties_2004} through an iterative search of parameter space that exhibits switch behavior. The initial markings in Table \ref{tab:transition_markings} were the same as those used by Heiner et al. and Kohn et al.. The resulting transition to CPN showed correspondence to Heiner et al. allowing us to continue with the implementation of ILK expression as part of the network.\\
\begin{table}[h]
\scriptsize
\centering
\begin{tabular}{|c|c|c|c|}
\hline
\textbf{Transition} & \textbf{Final k-value} & \textbf{Transition} & \textbf{Final k-value} \\
\hline
k1 & 0.1929 & k19 & 0.1392\\
k2 & 0.0007 & k19 & 0.1392\\
k3, k21 & 0.0148 & k20 & 0.2144\\
k4, k22 & 1.6732 & ILK\_tr1 & 1.0000\\
k5, k29 & 0.2681 & ILK\_tr2 & 1.0000\\
k6, k30 & 0.0809 & ILK\_deg & 3.0000\\
k12, k15 & 1.5478 & FL1 & 0.4181 \\
k13, k16 & 0.0416 & FL2 & 0.4181\\
k14, k17 & 0.0226 & FL3 & 0.4181 \\
k18 & 0.4738 & pmTOR\_deg & 3.0000 \\
pAkt\_deg & 0.2000 & &\\
\hline
\end{tabular}
\caption{Firing Rate values set (k-values set) for each transition MassActionFunction.}
\label{tab:reaction_rates}
\end{table}


\begin{table}[h]
\scriptsize
\centering
\begin{tabular}{|c|c|}
\hline
\textbf{Place} & \textbf{Marking} \\
\hline
VHL & 10 \\
PHD & 10 \\
ARNT & 5 \\
HRE & 1 \\
O2 & 0.1-0.9 \\
\hline
\end{tabular}
\caption{Initial markings used. The value of oxygen varies from 0.1 to 0.9 for the simulations. Places that had an initial marking of zero are left out from the table.}
\label{tab:transition_markings}
\end{table}
\subsection{ILK expression}
 The first step in establishing ILK expression was to define the ILK gene. As the original network already includes a general HRE, this HRE in the context of this project was treated as the binding site on the gene encoding the ILK protein. While not applied in a Petri net, transcription of an HRE has already been modelled \cite{kohn_properties_2004,yu_pathway_2007} as a sequence of intermediate reactions forming mRNA. The intermediate reactions in this model serve solely as a time delay for expression and do not represent any specific biological processes. Similarly, Kohn et al. implemented mRNA to model a VHL-dependent feedback loop based on the concentration of HRE mRNA.\\
To implement a similar expression in the context of Petri nets, two parallel transitions were introduced which represented the transcription of the ILK HRE to ILK coming from the two transcription factor gene complexes (HIF:ARNT:HRE and HIFOH:ARNT:HRE). Because this reaction is not a conversion of gene to its expressed protein a reverse arc was added coming from the transition back to the gene complex. While the intermediate steps implemented by Kohn et al. are useful for studying the complex dynamic behavior of their network. But a time delay was deemed unnecessary for this model, as the focus is primarily on the steady-state values of the system. Therefore, only one transition representing transcription was included for each HIF:HRE complex, termed ``ILK\_tr1" and ``ILK\_tr2". An additional transition termed ``ILK$\_$deg" was further added representing the protein's degradation. Such a degradation term allows for eventually reaching a steady state concentration value and prevents the resulting ILK protein from increasing indefinitely. 
Given that the complexity of gene expression was not central to this study, the formation of an ILK protein was simplified as a product of one ``transcription" transition.
A final addition to the network was a change in the k-value set (Table \ref{tab:reaction_rates} Intermediate k-value and Final k-value). Kohn et al \cite{kohn_properties_2004} obtained three different sets of k-values with the third one tethered for their extended network. This set of values was considered a more accurate choice, given that the network also implements transcription and a feedback loop. A particularly significant change in these values is the increase in the k12 value, which appears to compensate for the additional input of tokens towards HIF due to the feedback loop.\\

The next step in the modeling process was to implement an ILK feedback loop towards HIF, as described by Chou et al. \cite{chou2015novel} (see Fig.\ref{fig:ILK-signalling}). Two additional places were introduced into the network: p-473S-Akt and p-mTOR, connecting ILK with HIF through simple unidirectional transitions as a starting point. These transitions were termed FL1, FL2 and FL3. Since there was no specific reference for choosing k-values for these transitions, an average value of the k-values from the core HIF network was used. Applying average k-values provides a consistent starting point, so the effect of the ILK feedback loop on HIF signaling could be explored without guessing random values. A consequence of this choice is that it might oversimplify how the ILK feedback loop works. However, it allows for the detection of general patterns in the network. \\
To prevent the simplicity of the structure resulting in the net converging to the same steady-state value, additional reverse arcs were introduced, similar to the design of the transcription transition. Moreover, degradation term is introduced to represent the natural degradation of proteins and it acts as an interface for the biological system. Therefore, degradation terms (pmTOR$\_$deg and pAkt$\_$deg) for each feedback loop protein were introduced. Initially the choice for these terms was 0.2 as chosen for the HIF mRNA degradation term in the feedback loop from Kohn et al.. However, the value of pmTOR$\_$deg was increased to ensure lower steady-state concentration relative to HIF as seen on Figure \ref{fig:western-blot}.  This resulted in the final network for this case study.

\begin{figure}[h!]
    \centering
    \includegraphics[width=0.5\textwidth]{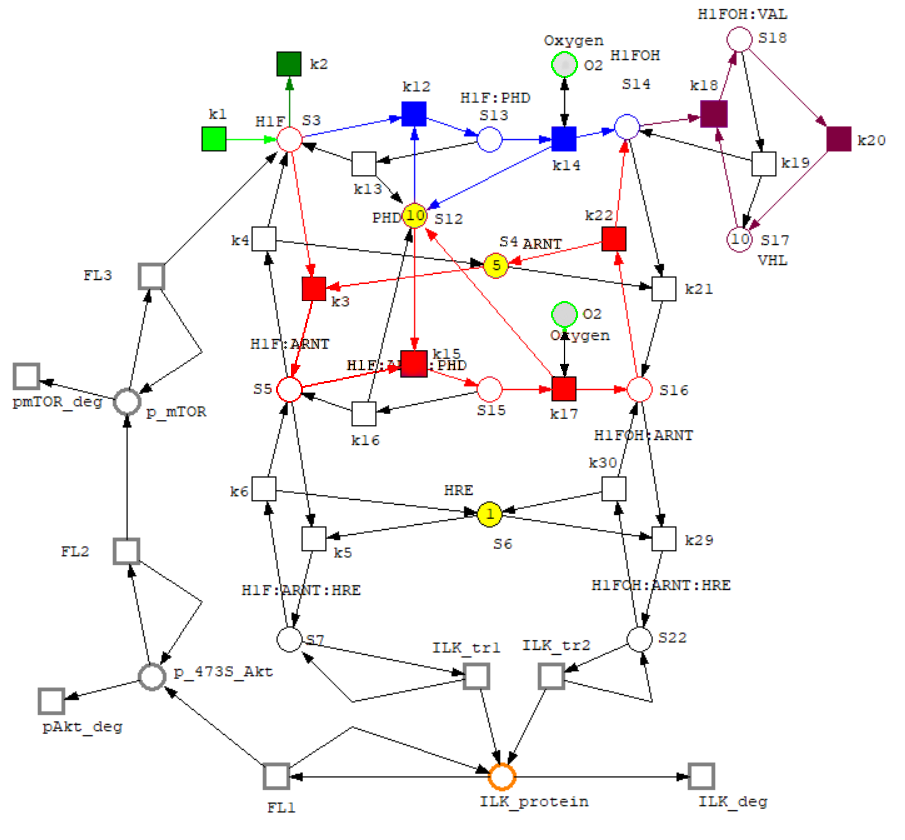}
    \caption{HIF-ILK feedback Petri net. Similar to Figure \ref{fig:HIFdiagram}, the green pathway is oxygen-independent, while the blue and red pathways are oxygen-dependent, all contributing to HIF transcription factor degradation.}
    \label{fig:HIFCPN}
\end{figure}

Petri model developed in this project were constructed using the Snoopy environment. The final model is shown in Figure \ref{fig:HIFCPN}. 

\section{\uppercase{Analysis and Results}}
Initial simulations in Snoopy were conducted for different marking sets to capture the dynamic behavior of the network under varying oxygen conditions. These simulation results were then compared with Western blot data, which provided experimentally observed expression levels. Lastly, Charlie software was utilized for a structural analysis, exploring various properties of the networks at hand.
\subsection{Simulations using Snoopy}
\begin{figure}
    \centering
    \includegraphics[scale=0.2]{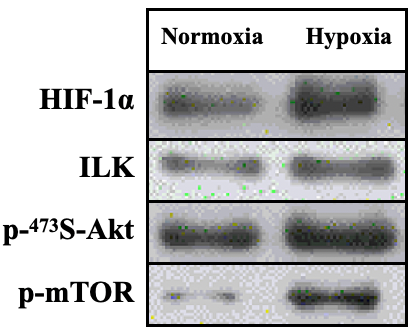}
    \caption{Western blot analysis of the
expression levels of endogenous HIF1-\(\alpha\), ILK, p-473S-Akt, and p-mTOR in MCF-7 cell line \cite{chou2015novel}.}
    \label{fig:western-blot}
\end{figure}

Simulations were performed for all values from 0.1 to 0.9 in increments of 0.1. This method allowed for the simulation of different degrees of hypoxia within the network. The behavior of HIF-1\(\alpha\) in relation to oxygen was observed. For simplicity, HIF-1\(\alpha\) is represented as HIF in the Petri net and will be referred to as HIF throughout this analysis. In addition, the behavior of the elements in the ILK feedback loop was also analyzed.\\
To verify the biological accuracy of the network, a comparison to an experimental Western blot was conducted for four proteins of interest: HIF1-\(\alpha\), ILK, p-473S-Akt, and p-mTOR. The results of this analysis are presented in Figure \ref{fig:western-blot}.  
These Western blots are sourced from a paper by Chou et al. that investigates the regulatory loop between HIF-1\(\alpha\) and ILK \cite{chou2015novel}. It can be observed that the intensity of the HIF-1\(\alpha\) and ILK bands is higher in hypoxia compared to normoxia, which aligns with expectations. The p-mTOR also shows much higher intensity under hypoxic conditions. For p-473S-Akt, the intensity is slightly higher in hypoxic conditions, but this change is not as noticeable as the other molecules. The results of Western blots suggest that the activation of Akt can be influenced by multiple signaling pathways, not solely by hypoxia, which might mask any hypoxia-specific changes.

\begin{figure}[h!]
    \centering
    \begin{subfigure}[b]{0.25\textwidth}
        \centering
        \includegraphics[width=0.9\textwidth]{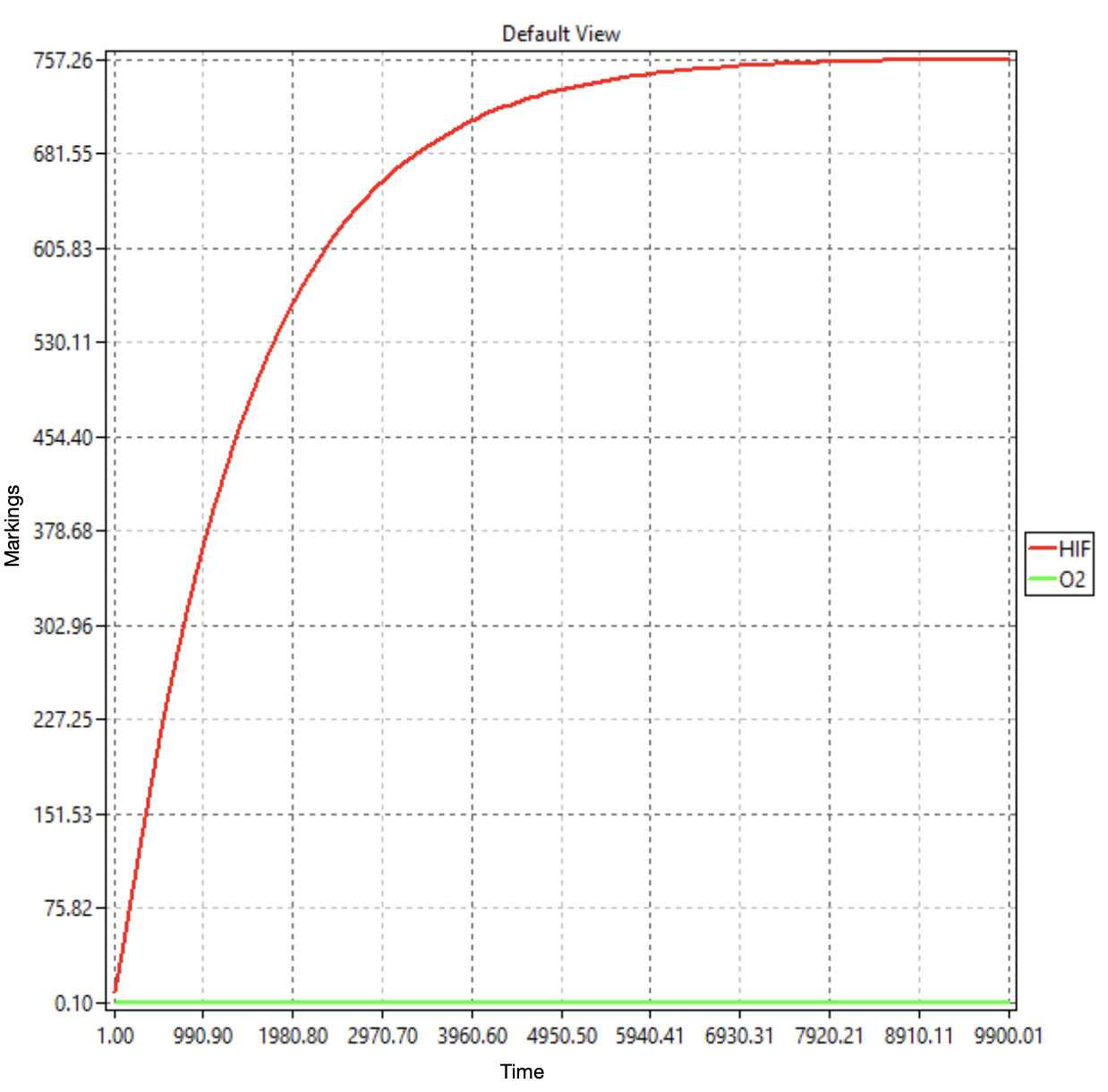}
        \caption{Initial \(O_2\) marking = 0.1.}
        \label{fig:final-low-o2-HIF}
    \end{subfigure}%
    \begin{subfigure}[b]{0.25\textwidth}
        \centering
        \includegraphics[width=0.83\textwidth,trim=0 0 18mm 0,clip=true]{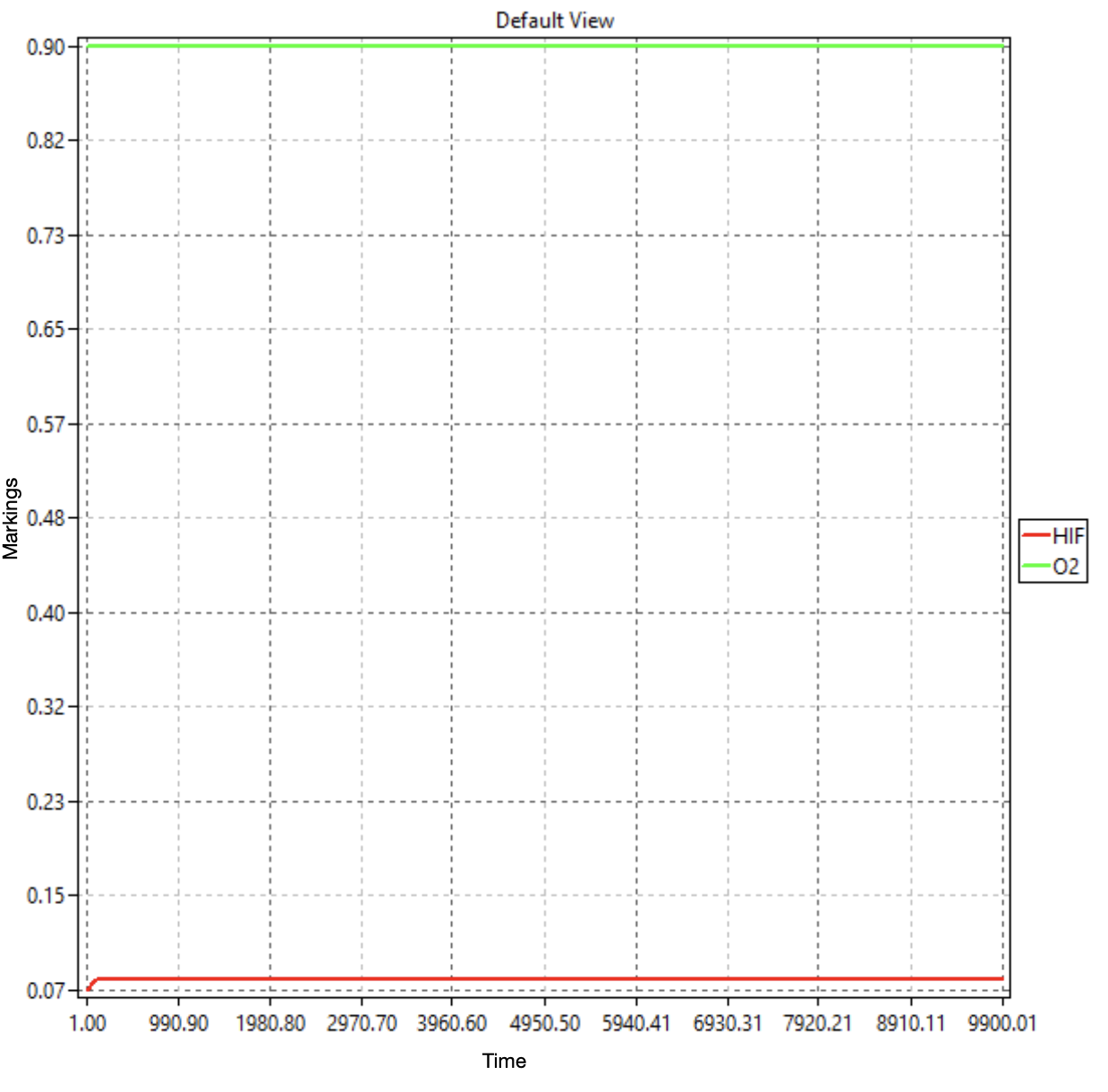}
        \caption{Initial \(O_2\) marking = 0.9.}
        \label{fig:final-high-o2-HIF}
    \end{subfigure}
    \caption{Snoopy simulation of the final Petri net, with a time range set to 10,000.}
    \label{fig:final-snoopy-HIF}
\end{figure}

The responses of HIF to hypoxia and normoxia for the final Petri net are illustrated in Figure \ref{fig:final-snoopy-HIF}. HIF has a steady-state expression level of 757.26 under hypoxia condition. The level of HIF under normoxia conditions stays around 0.07. 

\begin{figure}[h!]
    \centering
    \begin{subfigure}[b]{0.25\textwidth}
        \centering
        \includegraphics[width=\textwidth]{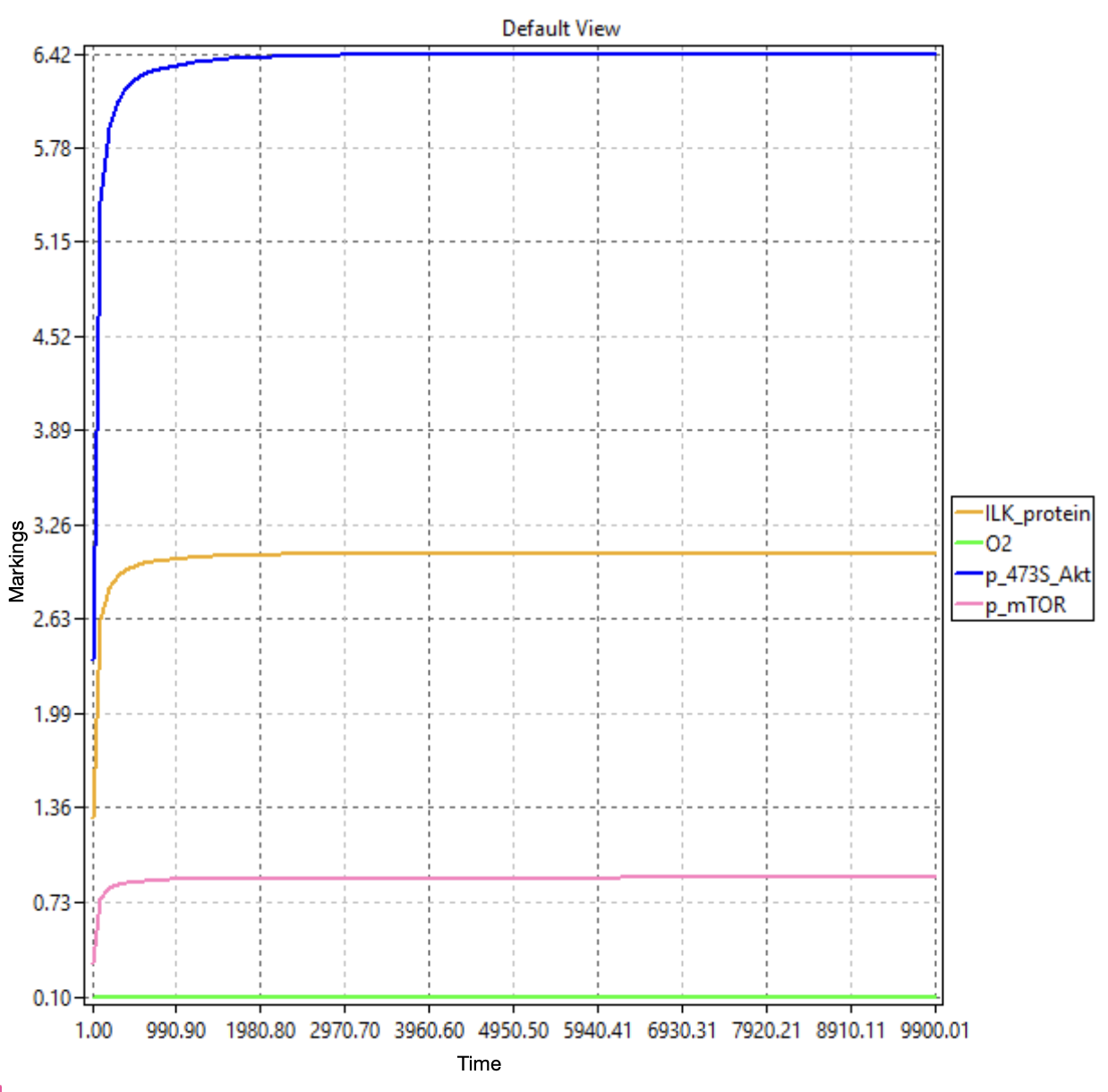}
        \caption{Initial \(O_2\) marking = 0.1.}
        \label{fig:final-low-o2_ILK}
    \end{subfigure}%
    \begin{subfigure}[b]{0.25\textwidth}
        \centering
        \includegraphics[width=0.83\textwidth,trim=0 0 32mm 0,clip=true]{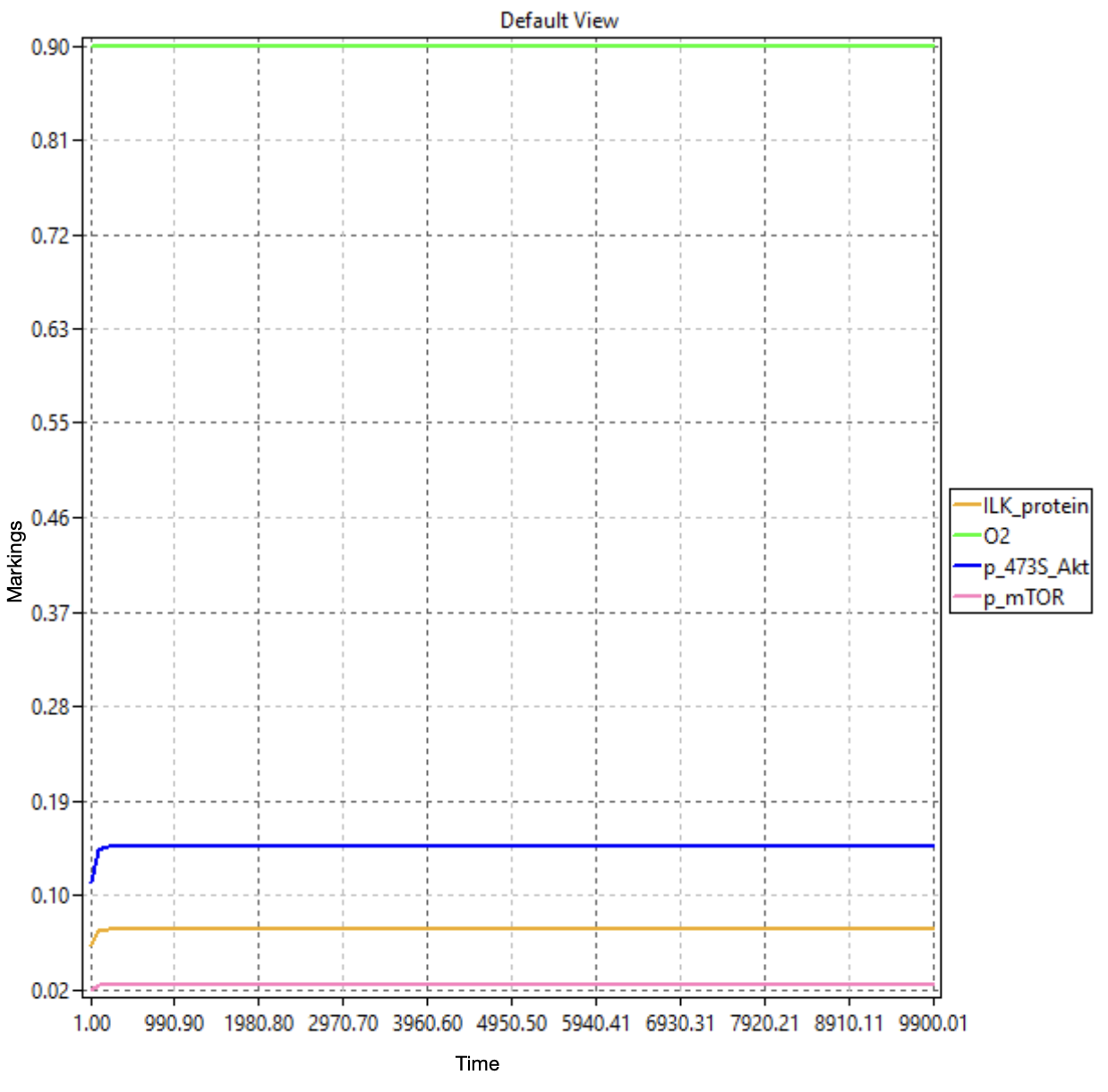}
        \caption{Initial \(O_2\) marking = 0.9.}
        \label{fig:final-high-o2-ILK}
    \end{subfigure}
    \caption{Snoopy simulation of the final Petri net for the ILK feedback loop, with a time range set to 10,000.}
    \label{fig:final-snoopy-ilk}
\end{figure}

The simulation of hypoxia and normoxia on the proteins of the ILK feedback loop are displayed in Figure \ref{fig:final-snoopy-ilk}. In general, the proteins are expressed more in hypoxia condition than in normoxia condition. The protein expression levels are consistent with the Western blot results shown in Figure \ref{fig:western-blot}. Specifically, p-473S-Akt is the most highly expressed, followed by the ILK protein and p-mTOR is the least expressed. This expression pattern was achieved by adjusting the degradation transitions to simulate the observed behavior. 

\subsection{Steady State Comparison}\label{steadystate}
\begin{figure}[h!]
    \centering
    \begin{subfigure}[b]{0.25\textwidth}
        \includegraphics[width=1\textwidth]{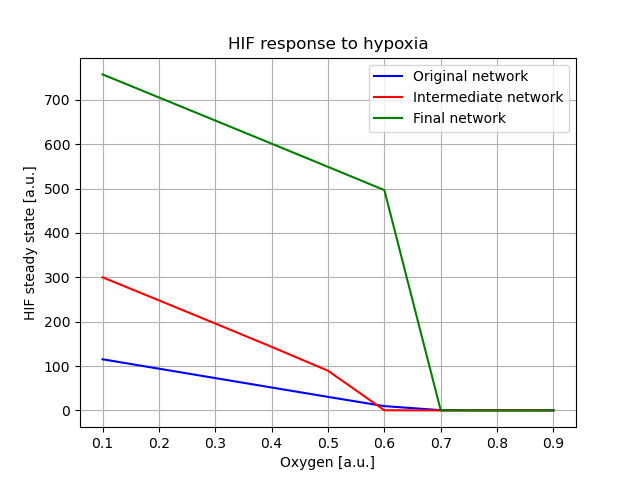}
        \caption{}
        \label{fig:HIF_steady_state}
    \end{subfigure}%
    \begin{subfigure}[b]{0.25\textwidth}
        \includegraphics[width=1\textwidth]{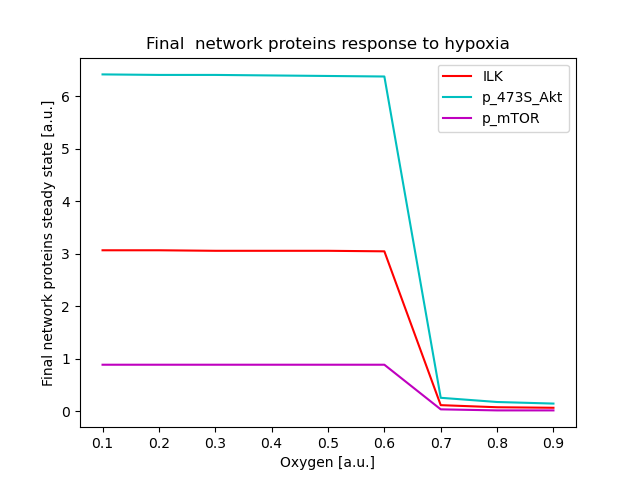}
        \caption{}
        \label{fig:steady-state-ilk}
    \end{subfigure}
    \caption{(a)Steady-state graph of HIF response to varying oxygen levels. Original network: the net incorporating the Petri net from Heiner et al.; Intermediate network: adding a complex feedback loop; Final network: further adding degradation processes. (b)Steady state graph of proteins: ILK, p-473S-Akt and p-mTOR, for the final network.}
\end{figure}

A network reaches a steady state when it achieves a stable equilibrium, these values are recorded and analysed in this section. Figure \ref{fig:HIF_steady_state} compares the steady state values of HIF in response to varying oxygen levels ranging from 0.1 to 0.9, with increments of 0.1. In the final network, HIF levels remain higher at lower oxygen levels compared to the original and intermediate networks. The HIF value drops significantly around oxygen level of 0.6, indicating a switch-like behavior. After which the HIF level stays constant at zero.

Figure \ref{fig:steady-state-ilk} depicts the response of proteins in the ILK pathway (ILK, p-473S-Akt, and p-mTOR) to varying oxygen levels within the final network. In the final network, the steady state values of the proteins drop between oxygen level of 0.6 and 0.7.

\subsection{Analysis using Charlie}
In this section, analysis results of our Petri net using Charlie\cite{heiner2015charlie} are displayed. Charlie provides a qualitative analysis of the structural and behavioral properties of the Petri nets. The conducted analysis aims to elucidate the fundamental characteristics and dynamic behaviors of biological networks. However, Charlie only accepts discrete Petri net. Therefore, we converted our continuous Petri net into a discrete Petri net for the analysis. The results derived from the structural analysis properties can be found in Table \ref{table:structural-properties}.
\begin{table}[h!]
\scriptsize
\centering
\begin{tabular}{p{0.04\textwidth}p{0.02\textwidth}p{0.08\textwidth}p{0.25\textwidth}}
\hline
\textbf{Property} & \textbf{Final} & \textbf{Full name} & \textbf{Definition}\\ \hline
PUR & \cellcolor{red!50}F & pure & No two places are connected in both directions.\\ 
ORD & \cellcolor{green!50}T & ordinary & The weight of all arcs equals to 1.\\ 
HOM & \cellcolor{green!50}T & homogeneous & All outgoing arcs of a place have the same multiplicity.\\ 
CON & \cellcolor{green!50}T & connected & For every two places, there exists an undirected path. \\ 
SC & \cellcolor{red!50}F & strongly connected & For every two places, there exists a directed path. \\ 
NBM & \cellcolor{green!50}T & non-blocking multiplicity & The minimum weight of incoming arcs of a place is greater or equal to the maximum weight of outgoing arcs of that place.\\ 
CSV & \cellcolor{red!50}F & conservative & Every transition adds the same number of tokens to its post-places as it subtracts from its pre-places.
\\ 
SCF & \cellcolor{red!50}F & structurally conflict free & There are no two transitions sharing a pre-place.\\ 
FT0 & \cellcolor{red!50}F & & Every transition has a pre-place.\\ 
TF0 & \cellcolor{red!50}F & & Every transition has a post-place.\\ 
FP0 & \cellcolor{green!50}T & & Every place has a pre-transition.\\ 
PF0 & \cellcolor{green!50}T & & Every place has a post-transition.\\
RKTH & \cellcolor{red!50}F & rank theorem & The rank of the incidence matrix of the Petri net equals to $\left| SCCS \right|-1$.\\ 
NC & \cellcolor{yellow!50}nES &net class& nES: not extended simple.\\
\hline
\end{tabular}
\caption{Structural properties of final Petri Net calculated using Charlie. "T" denotes true for possessing the property, while "F" denotes false, indicating the property is absent. Yellow indicates information beyond a binary classification.}
\label{table:structural-properties}
\end{table}

Properties like ORD (ordinary), HOM (homogeneous), CON (conservative), NBM (non-blocking), FP0 (every place has a pre-transition), and PF0 (every place has a post-transition) are present, indicating that the net maintains uniformity, conservation, and avoids blocking in its structure. This reflects a consistent and balanced distribution of tokens across places and transitions. However, properties such as PUR (pure), SC (strongly connected), CSV (conservative), SCF (structurally conflict free), FT0 (Every transition has a pre-place), TF0 (Every transition has a post-place), and RKTH (rank theorem) are absent. The absence of these properties suggests that the net contains self-loops, lacks a strongly connected component, has transition without pre-place or post-place, has structural conflicts, and doesn't satisfy the rank theorem conditions. These absences highlight areas where the net's structure may not fully adhere to certain biological consistencies or principles. \\

The analysis of the behavioral properties of the Petri nets under hypoxia and normoxia conditions reveals that the behavioral properties change between oxygen levels. Key properties such as k-Boundedness (k-B) and Structural Boundedness (SB) are absent across all conditions, indicating that the nets are unbounded and have infinite state spaces, making reachability graph visualization impractical. The Siphon-Trap Property (STP) is present under normoxia but absent under hypoxia, suggesting enhanced stability and liveness in normoxic conditions. Reversibility (REV) is false, indicating that the net cannot revert to their initial state from any reachable state. Lastly, the absence of dead states (Dst) under normoxia suggests that normoxic conditions help maintain continuous operation by avoiding states with no possible transitions. For the hypoxic conditions this property was not discovered.\\

The P-invariants are presented in Table \ref{table:p-invariants}. Each P-invariant represents a set of places in the Petri net where the total number of tokens remains constant, reflecting the conservation of certain resources or substances within the network. In our network, the amount of oxygen ($O_2$) remains constant throughout the network's processes. This constancy is based on the assumption that the system operates in an environment where oxygen levels are maintained based on the original marking. Similarly, the total amount of PHD proteins is conserved across places $S_{12}$, $S_{13}$, and $S_{15}$, representing free PHD, PHD bound to HIF1-\(\alpha\), and PHD bound to the HIF1 complex (ARNT and HIF) respectively. This consistency is logical since the enzyme merely catalyzes the reaction without being consumed in the process. The ARNT is conserved in multiple places, including $S_4$, $S_5$, $S_7$, $S_{15}$, $S_{16}$, and $S_{22}$. These places again represent all the bound and unbound versions of this protein. As it is part of a transcription factor it again isn't consumed but simply just binds other elements in the Petri net. The VHL is conserved in places $S_{17}$ and $S_{18}$, indicating that the total amount of the VHL protein remains constant in these locations. Similarly, the HRE are conserved in places $S_6$, $S_7$, and $S_{22}$. HREs act as binding sites for transcription factors but are not depleted during the regulatory process. The conservation of specific proteins and elements aligns with biological expectations, where certain quantities must remain stable to maintain cellular functions under different conditions. \\

\begin{table}[h!]
\centering
\small
\begin{tabularx}{0.4\textwidth}{XXX}
\hline
\textbf{Invariant} & \textbf{P-invariant name} & \textbf{Places}\\ \hline
1& Oxygen& $O_2$\\
2& $PHD_{total}$& $S_{12}$,$S_{13}$,$S_{15}$\\ 
3& $ARNT_{total}$& $S_4$,$S_5$,$S_7$,$S_{15}$,$S_{16}$,$S_{22}$ \\
4& $VHL_{total}$& $S_{17}$,$S_{18}$\\
5& $HRE_{total}$& $S_6$,$S_7$,$S_{22}$\\\hline
\end{tabularx}
\caption{P-invariants of the Petri net.}
\label{table:p-invariants}
\end{table}
\noindent

The non-trivial T-invariants were presented in Table \ref{table:t-invariant for HIF Pathways and ILK Pathways}. Each T-invariant represents a sequence of transitions in the Petri net that, when executed, return the network to its original state. Biologically, T-invariants represent consistent, repeatable behaviors. They identify sub-networks that either return to a given state after a series of reactions or maintain a steady state through continuous reactions. In the final network four T-invariants were identified. The first two represent the oxygen-dependant degradation of HIF mediated by PHD. The third and fourth T-invariants illustrate similar oxygen-dependent degradation pathways of HIF. However, in these T-invariants, the HIF does not come from the inflow into the system represented by the Petri net, but from the positive feedback resulting from the ILK loop. These four T-invariants effectively and intuitively show the input-output behavior of the network.  

\begin{table}[h!]
\centering
\small
\begin{tabularx}{0.4\textwidth}{p{1cm}X}
\hline
\textbf{Invariant} & \textbf{Transitions}\\ \hline
\hline
$1$& $k_1$, $k_{12}$, $k_{14}$, $k_{18}$, $k_{20}$\\
$2$& $k_1$, $k_3$, $k_{15}$, $k_{17}$, $k_{18}$, $k_{20}$, $k_{22}$\\
$3$& $k_{12}$, $k_{14}$, $k_{18}$, $k_{20}$, $FL3$\\
$4$& $k_{3}$, $k_{15}$, $k_{17}$, $k_{18}$, $k_{20}$, $k_{22}$, $FL3$\\
\hline
\end{tabularx}
\caption{T-invariants of the Petri net.}
\label{table:t-invariant for HIF Pathways and ILK Pathways}
\end{table}

\section{\uppercase{Conclusion}}
This paper aimed to investigate how variations in oxygen levels affect the expression of HIF and ILK pathways through a continuous Petri net. The analysis with the Snoopy simulations show that hypoxic conditions result in higher expression of HIF and ILK pathway compared to normoxic conditions. The final network achieves a steady-state equilibrium, with expression levels corresponding to the Western blot used for validation, thereby confirming that the final network functions as expected. When examining the steady-state graphs of the final network, the expression levels of the HIF and ILK pathways switch from hypoxic to normoxic conditions between an oxygen concentration of 0.6 and 0.7. Additionally, the structural, behavioral, T-invariant, and P-invariant analyses provide deeper insights into the network's properties.\\ 
Overall, this study provides a robust model for understanding the dynamics of these pathways under different oxygen conditions, which can be crucial for further research in cancer biology and the development of therapeutic strategies targeting hypoxia-induced cancer behavior.\\
To broaden the scope and enhance the robustness of the current Petri net model, future research should integrate more experimental data and focus on additional breast cancer cell lines, such as triple-negative and HER2-positive. This expansion would improve our understanding of varying hypoxia responses and ILK inhibitor sensitivities across different subtypes, potentially uncovering novel therapeutic targets within the HIF-ILK pathway. Furthermore, incorporating the ILK inhibitor T315, as studied by Chou et al. \cite{chou2015novel}, into the model could provide valuable insights into its ability to disrupt the HIF-ILK loop. This integration would enhance the model's accuracy and offer implications for more effective therapeutic strategies against hypoxia-driven cancer progression.

\bibliographystyle{apalike}
{\small
\bibliography{Paper}}

\end{document}